\documentclass[12pt]{article}
\usepackage{cite}
\usepackage{amsmath,amsfonts,amssymb}
\usepackage{pstricks,pst-node}
\usepackage[small,bf,hang]{caption2}
\usepackage[dvips]{graphicx,epsfig}
\usepackage[vcentermath,enableskew]{youngtab}

\def\hybrid{
        \topmargin -20pt
        \oddsidemargin 0pt
        \headheight 0pt \headsep 0pt
        \textwidth 6.25in       
        \textheight 9.5in       
        \marginparwidth .875in
        \parskip 5pt plus 1pt   \jot = 1.5ex}

\hybrid

 \csname
@addtoreset\endcsname{equation}{section}

\newcommand{\bein}[3]{{#1}_{#2}^{\hspace{0.5em}{#3}}}

\def\moth{\mathsurround=0pt}
\newdimen\zo \zo=0pt

\def\tick{\leaders\hrule height 0.5ex depth 0pt \hskip 0.5pt}
\def\upboxfill{$\moth \setbox\zo\hbox{\tick}%
  \hskip 3pt\hbox to 0pt{$\tick$\hss}\hrulefill \hbox to 7.5pt{$\tick$\hss}$}
\def\underbox#1{\offinterlineskip{\mathord{\mathop{\vtop{\moth\ialign{##\crcr
      $\hfil\displaystyle{#1}\hfil$\crcr\noalign{}
      {\upboxfill}\crcr\noalign{}}}}\limits}}}
\def\dtick{\leaders\hrule height .34pt depth 0.5ex \hskip 0.5pt}
\def\downboxfill{$\moth \setbox\zo\hbox{\dtick}%
  \hskip 2pt\hbox to 0pt{$\dtick$\hss}\hrulefill \hbox to 2pt{$\dtick$\hss}$}

\def\undersym#1{\underbox{{}#1}}

\def\mso{\mathfrak{so}}

\def\mhso{\mathfrak{ho}}

\def\bec{\begin{center}}
\def\ec{\end{center}}
\def\a{\alpha}

\def\c{\gamma} 

\def\d{\delta} 
\def\D{\Delta}
\def\e{\epsilon}

\def\L{\Lambda}
\def\m{\mu}
\def\n{\nu}
\def\r{\rho}

\def\om{\omega}

\def\cB{{\cal B}}

\def\cF{{\cal F}}
\def\cO{{\cal O}}
\def\cA{{\cal A}}

\def\cU{{\cal U}}

\def\cI{{\cal I}}

\def\cW{{\cal W}}

\def\cA{{\cal A}}

\def\cO{{\cal O}}

\def\del{\partial}

\let\la=\label

\def\nn{\nonumber}

\def\tr{{\rm tr}}
\def\Tr{{\rm Tr}}

\def\be{\begin{equation}}
\def\ee{\end{equation}}
\def\bea{\begin{eqnarray}}
\def\eea{\end{eqnarray}}
\def\ba{\begin{array}}
\def\ea{\end{array}}

\begin{document}

\begin{titlepage}
\begin{center}

\hfill UG-08-06 \\

\vskip 1.5cm

{\Large \bf Higher-spin dynamics and Chern-Simons theories
\\[0.2cm]}

\vskip 1.5cm

{\bf Johan Engquist\footnotemark
\footnotetext{j.p.engquist@fys.uio.no} and Olaf Hohm\footnotemark}
\footnotetext{o.hohm@rug.nl}

\vspace{0.8cm} $^1${\em } {\em Department of Physics, University of
Oslo}  \\ P.O. Box 1048 Blindern, N-0316 Oslo, Norway\ \vspace{.5cm}

$^2${\em} {\em Centre for Theoretical Physics, University of
Groningen} \\ Nijenborgh 4, 9747 AG Groningen, The Netherlands \
\vspace{.5cm}

\vskip 0.8cm

\end{center}

\vskip 1cm

\begin{center} {\bf ABSTRACT}\\[3ex]

\begin{minipage}{13cm}
We review the construction of consistent higher-spin theories
  based on Chern-Simons actions. To this end we first introduce the
  required higher-spin algebras and discuss curvature and torsion
  tensors in an unconstrained way. Finally we perform a
  perturbative analysis of the Chern-Simons theory in $D=5$ for
  a non-maximally symmetric $AdS_4$ background and obtain the required
  four-dimensional Fr\o nsdal equations in the compensator formulation.

\end{minipage}

\vskip 3cm

\textit{Presented at the RTN workshop `ForcesUniverse', Valencia,
Spain, October 2007.}

\end{center}

\noindent

\vfill

April 2008

\end{titlepage}

\section{Introduction}\setcounter{equation}{0}

Higher spin (HS) theories have attracted increasing interest not
only due to their prominent appearance in string theory, but also as
a challenging problem on its own. In fact, until today the problem
of finding consistently interacting HS theories (e.g.~coupled to
gravity) remains without a satisfactory solution. The main obstacle
of formulating interactions for, say, massless HS fields is due to
the fact that these need to permit a HS gauge symmetry in order to
eliminate the longitudinal degrees of freedom. In particular, for
fields with spin higher than $3/2$ this seems to rule out the
possibility of consistent couplings to gravity
\cite{Aragone:1979bm,Sorokin:2004ie}.

One approach to circumvent the no-go theorems for gravity-HS
couplings has been pioneered by Vasiliev
\cite{Vasiliev:1990en,Vasiliev:2003ev}. It is based on the gauging
of certain \textit{infinite-dimensional} HS algebras in a similar
spirit as supergravity theories can be viewed as gauge theories of
(AdS-)supergroups. However, the actual formulation of the dynamics
is a severe problem since any standard coupling like the
Einstein-Hilbert term singles out part of the gauge field (as the
vielbein) and therefore breaks the symmetry. A manifestly invariant
formulation for HS gauge theories is given by the so-called unfolded
formulation \cite{Vasiliev:2005zu}, which is, however, only defined
at the level of the equations of motion. In contrast, a fully HS
invariant action principle was unknown, with the only exception
being the Chern-Simons theory in $D=3$ constructed in
\cite{Blencowe:1988gj}.

Here we are going to review \cite{Engquist:2007kz}, in which the
construction of Chern-Simons theories based on HS algebras has been
extended to generic odd dimensions. While the three-dimensional
theory is purely topological, remarkably, this is not so in higher
dimensions. To be more precise, around maximally symmetric
backgrounds there are still no non-trivial excitations, but around
less symmetric solutions there are \cite{Chamseddine:1990gk}. We
will see explicitly that linearizing the Chern-Simons theory in
$D=5$ around an $AdS_4\times S^1$ solution, gives precisely rise to
the required Fr\o nsdal equations for free HS fields on $AdS_4$. To
this end we will first introduce in sec.~2 a formulation of HS
theories in terms of frame-like fields based on HS algebras.

\section{Higher-spin algebra and geometry}
The HS gauge algebras are infinite-dimensional extensions of the AdS
algebra $\mathfrak{so}(D-1,2)$ in $D$ dimensions. The latter is
spanned by $M_{AB}$, $A,B=1,\ldots,D+1$, satisfying the standard
algebra
 \begin{equation}
  [M_{AB},M_{CD}] \ = \
  \eta_{BC}M_{AD}-\eta_{AC}M_{BD}-\eta_{BD}M_{AC}+\eta_{AD}M_{BC}\;,
 \end{equation}
where $\eta_{AB}$ denotes the $\mso(D-1,2)$ invariant metric. The HS
algebra $\mhso(D-1,2)$ is in turn given by the enveloping algebra
$\cU(\mso(D-1,2))$ of the AdS group, divided by a certain ideal,
 \bea\label{hsalgebra}
  \mhso(D-1,2) \ = \ \cU(\mso(D-1,2))/ \cI\;.
 \eea
More precisely, the enveloping algebra is spanned by all polynomials
in $M_{AB}$, while modding out some ideal reduces these to a certain
subclass corresponding to restricted Young tableaux. A minimal
choice then leads to HS generators transforming as tensors in two
row Young tableau under $\mso(D-1,2)$:
 \begin{equation}\label{generators}
  Q_{A(s-1),B(s-1)}: \qquad \underbrace{\yng(5,5)\cdots\yng(5,5)}_{s-1}\
  ~~  ,
 \end{equation}
where we employed the notation $A(s)=A_1\cdots A_s$. As we will see
below, these give precisely rise to all states which carry integer
spin. Other choices would contain mixed HS states corresponding to
fields in mixed Young tableaux representations, which do appear in
dimensions higher than four and in particular in string theory. That
$\mhso(D-1,2)$ is a consistent Lie algebra can be proved, for
instance, by means of an explicit realization in terms of vector
oscillators. However, the Lie brackets are not known in a closed
form, and so we will here focus only on the lowest-order terms,
which is sufficient for our linearized analysis below. Explicitly,
this means that we focus on the Lie brackets
 \begin{equation}\label{brackets}
  [M_{AB},Q_{C(s-1),D(s-1)}] \ = \ -4(s-1)\eta\undersym{{}_{A\langle
  C_{s-1}}Q_{|B|}}\hspace{0cm}{}_{C(s-2),D(s-1)\rangle} \;,
 \end{equation}
which are fixed by representation theory. Here, brackets
$\langle\hspace{0.2em}\rangle$ denote Young projection according to
(\ref{generators}).

Let us now examine the gauge theory based on a HS algebra in more
detail. In order to read off the `physical' fields contained in a
gauge connection based on $\mhso(D-1,2)$, we need to split the
generators (\ref{generators}) into Lorentz covariant tensors,
i.e.~we decompose the generators into $Q_{a(s-1),b(t)}$ for $0\leq
t\leq s-1$ with Lorentz indices $a,b,\ldots =1,\ldots,D$,

\vspace{-3.0cm}
  \setlength{\unitlength}{0.8cm}
  \begin{picture}(10,5)
  \put(2.45,-0.13){$Q_{a(s-1),b(t)}$:}\put(13.0,-0.13){.}
  \put(7.55,-0.13){$\overbrace{\yng(8,5)}^{s-1}$}
    \put(7.55,-0.45){$\underbrace{\phantom{\yng(5)}}_{t}$}
  \end{picture}

  \vspace{1cm} \

\vspace{.0cm} Next we introduce the Lie algebra valued gauge field,
 \begin{equation}\label{connection}
  {\cal A}_{\mu}\ = \ \bein{\bar{e}}{\mu}{a}P_a +
  \frac{1}{2}\bein{\bar{\omega}}{\mu}{ab}M_{ab} + \sum_{s=3}^\infty\cW_\m^{(s)}\;,
 \end{equation}
where $\bein{\bar{e}}{\mu}{a}$ and $\bein{\bar{\omega}}{\mu}{ab}$
denote vielbein and spin connection of the background geometry,
corresponding to the translation and Lorentz generators,
respectively. The spin-$s$ contribution is given by
  \bea
   \cW_\m^{(s)}& =&
   \frac{1}{(s-1)!}e_\mu{}^{a(s-1)}Q_{a(s-1)}+\sum_{t=1}^{s-1}\frac{s-t}{s!t!}\om_\mu{}^{a(s-1),b(t)}Q_{a(s-1),b(t)}\ .
  \eea
The fields $\om_\mu{}^{a(s-1),b(t)}$ will be interpreted in the
following as HS connections, while $e_\m{}^{a(s-1)}$ corresponds to
the physical spin-$s$ field, or the generalized vielbein. The
non-abelian curvature derived from (\ref{brackets}) decomposes into
 \begin{eqnarray}
   \la{rt99}
   \cF_{\m\n}^{(s)}\ = \ \sum_{t=0}^{s-2}\frac{s-t}{s!t!}T_{\m\n}{}^{a(s-1),b(t)}Q_{a(s-1),b(t)}
   +\frac{1}{s!(s-1)!}R_{\mu\nu}{}^{a(s-1),b(s-1)}Q_{a(s-1),b(s-1)}\
   ,
 \end{eqnarray}
whose explicit form is given by
 \begin{eqnarray} \label{strength}
    \begin{split}
        R_{\m\n}{}^{a(s-1),b(s-1)}& \ = \ \bar
        D_{\m}\om_\n{}^{a(s-1),b(s-1)}+2(s-1) \L\om_\m{}^{\langle
        a(s-1),b(s-2)} \bar e_\n{}^{b_{s-1}\rangle}- (\mu \leftrightarrow \n)\ , \\
        T_{\m\n}{}^{a(s-1),b(t)}& \ = \ \bar
        D_{\m}\om_\n{}^{a(s-1),b(t)}+\om_\m{}^{
        a(s-1),b(t)c}\bar e_{\n c}\\ &\qquad +t(s-t+1)\L\om_\m{}^{\langle
        a(s-1),b(t-1)}\bar e_\n{}^{b_t\rangle}- (\mu \leftrightarrow \n)\ ,
    \end{split}
\end{eqnarray}
where $\bar{D}_{\mu}$ denotes the background Lorentz covariant
derivative. The components (\ref{strength}) will be interpreted as
HS Riemann and torsion tensors. Finally we give the non-abelian HS
gauge transformations, $\delta {\cal A}_{\mu} = D_{\mu}\epsilon =
\partial_{\mu}\epsilon+[{\cal A}_{\mu},\epsilon]$, under which the
linearized curvatures (\ref{strength}) are invariant: \bea
\la{gaugetr}
    \begin{split}
        \d_\e e_\m{}^{a(s-1)} & \ =\  \bar D_\m \e^{a(s-1)} -\e^{
        a(s-1),c}\bar e_{\m c}\ , \\
        \d_\e \om_\m{}^{a(s-1),b(t)} &\ = \ \bar D_{\m}\e^{a(s-1),b(t)}-\e^{
        a(s-1),b(t)c}\bar e_{\m c}-t(s-t+1)\L\e^{\langle
        a(s-1),b(t-1)}\bar e_\m{}^{b_t\rangle}\ .
    \end{split}
 \eea
The symmetry parameterized by $\epsilon^{a(s-1)}$ will give rise to
the physical HS symmetry. In addition, the symmetries given by
$\epsilon^{a(s-1),b(t)}$ for $t\geq 1$ act as St\"uckelberg shift
symmetries, that generalize the linearized Lorentz transformations
of Einstein gravity.

Before we proceed let us make contact with the Fr\o nsdal
formulation \cite{Fronsdal:1978rb}, in which HS fields are
represented by totally symmetric tensors $h_{\mu_1\ldots\mu_s}$ of
rank $s$, satisfying the field equations
 \bea\label{fronsdal}
  {\cal F}_{\mu_1\cdots \mu_s}\ = \ \square h_{\mu_1\cdots \mu_s}
  -s\nabla_{(\mu_1}\nabla\cdot
  h_{\mu_2\cdots\mu_s)}+\frac{s(s-1)}{2}\nabla_{(\mu_1}^{}
  \nabla_{\mu_2}^{}h_{\mu_3\cdots\mu_s)}^{\prime}+\cO (\Lambda) \ = \ 0\;,
 \eea
where $\nabla\cdot$ denotes the divergence and $h^{\prime}$ the
trace in the AdS metric. The relation between the HS fields
$e_{\mu}{}^{a(s-1)}$ encountered above and the Fr\o nsdal fields is
precisely analogous to the relation between vielbein and metric in
general relativity. In the latter, the torsion constraint allows to
solve for the spin connection in terms of the vielbein. After
gauge-fixing the Lorentz symmetry, only the symmetric part of the
vielbein survives, which then coincides with the ordinary metric
tensor. In the HS case, the torsion constraints we are going to
impose are given by
 \bea\label{constraint}
  T_{\m\n}{}^{a(s-1),b(t)} \ = \ 0 \ , \qquad 0\le t\le s-2\ \;.
 \eea
In order to solve this chain of constraints, it turns out to be
convenient to gauge-fix the St\"uckelberg shift symmetries in
(\ref{gaugetr}). The lowest component of the HS connection then
amounts to the totally symmetric Fr\o nsdal field
$h_{\mu_1\ldots\mu_s}\equiv e_{(\mu_1|\mu_2\ldots\mu_s)}$, where we
converted all indices into curved ones. The gauge symmetry
(\ref{gaugetr}) then reads
 \bea\label{gaugefixtr}
  \delta_{\epsilon}h_{\mu_1\ldots\mu_s} \ = \
  \nabla_{(\mu_1}\epsilon_{\mu_2\ldots \mu_{s})}\;,
 \eea
which is the HS symmetry that leaves (\ref{fronsdal}) invariant. The
first connection can be expressed in terms of derivatives of the
physical HS field, the second connection in terms of the first
connection, etc. In total, this yields a chain of connections, each
being expressible in terms of $t$ derivatives of the physical HS
field. Their explicit form is given by
 \begin{eqnarray}\label{explconn}
  \nn\om_{\m|\n(s-1),\r(t)}& = &
     \frac{s}{s-t}\nabla_{\langle\r_1}\cdots\nabla_{\r_t}h_{\n(s-1)\rangle\m}\\ \la{omeads9} &&
     +\sum_{k=1}^{[t/2]}\L^k\c_{k,t}g_{\langle\r_1\r_2}\cdots
     g_{\r_{2k-1}\r_{2k}}\nabla_{\r_{2k+1}}\cdots\nabla_{\r_t}h_{\n(s-1)\rangle\m}\
     ,
 \end{eqnarray}
where $g_{\mu\nu}$ is the AdS metric, and we refer to
\cite{Engquist:2007yk} for the coefficients $\c_{k,t}$.

Next we are going to explain in which way the free HS dynamics is
encoded in this geometrical formalism. First of all it is puzzling
how to obtain sensible second order field equations (the Fr\o nsdal
equations (\ref{fronsdal})), since the HS invariant Riemann tensor
is an $s$-derivative object in the physical HS field. However, in
\cite{Bekaert:2003az,Sagnotti:2005ns,Bekaert:2006ix} it has been
shown in flat space that the HS Einstein equation
--- stating vanishing of the HS Ricci tensor --- gives effectively
rise to second order equations through local integrations, in which
the `integrations constants' correspond to gauge degrees of freedom.
To explain this, let us focus on the first non-trivial case, namely
a spin-3 field on Minkowski space. Inserting (\ref{explconn}) into
the HS Riemann tensor (\ref{strength}) in the limit $\Lambda=0$ and
taking the trace yields \cite{Damour:1987vm}
 \bea\label{RieTr}
  R_{\mu\nu\hspace{0.3em}\rho\sigma},{}^{\lambda}{}_{\lambda}\
  \equiv \ ({\rm Ric})_{\mu\nu|\rho\sigma}\ = \ 2\partial_{[\mu}{\cal
    F}_{\nu]\rho\sigma} \ = \  0\;.
 \eea
Here ${\cal F}$ is the Fr\o nsdal operator defined in
(\ref{fronsdal}). Equation (\ref{RieTr}) shows that the Ricci tensor
is a curl, which can therefore be locally integrated by virtue of
the Poincar\'e lemma, resulting in ${\cal F}_{\mu\nu\rho} =
\partial_{\mu}\alpha_{\nu\rho}$. Since the right-hand side has to be
totally symmetric, this implies
$\alpha_{\nu\rho}=\partial_{\nu}\partial_{\rho}\alpha$, i.e.~in
total
  \bea\label{compensator}
  {\cal F}_{\mu\nu\rho} \ =\
  \partial_{\mu}\partial_{\nu}\partial_{\rho}\alpha\;.
 \eea
These are the so-called compensator equations
\cite{Francia:2002aa,Francia:2002pt}. They possess a larger symmetry
than the actual Fr\o nsdal equations ${\cal F}=0$. While the latter
are invariant only under so-called constrained transformations with
$\epsilon^{\prime}=\epsilon^{\mu}{}_{\mu}=0$, the compensator
equations are completely invariant under (\ref{gaugefixtr}) by
virtue of the shift transformation
$\delta_{\epsilon}\alpha=\epsilon^{\prime}$ on the compensator
$\alpha$. These shift symmetries can in turn be used to set
compensator to zero, and so one recovers precisely the spin-3 Fr\o
nsdal equations (\ref{fronsdal}). In other words, despite being of
higher derivative order, (\ref{RieTr}) correctly describes a
massless spin-3 field on flat space. It has been shown in
\cite{Engquist:2007yk} that this pattern generalizes to all HS
fields on AdS, i.e.~the HS Einstein equations as in (\ref{RieTr})
correctly account for the dynamics of massless HS fields on flat
space as well as AdS.

\section{Higher-spin Chern-Simons theories}
In this section we introduce Chern-Simons theories based on HS
algebras. To start with, we recall that Chern-Simons actions are
gauge-invariant and topological (in the sense that they do not
depend on a metric). In $D=5$, which is the generic situation we
will focus on the following, the action for a gauge connection
${\cal A}$ is given by
 \bea\label{CS}\la{cs}
  S=\int_{M_5}\Big{\langle} {\cal A}\wedge d{\cal
  A}\wedge
  d{\cal A}+\frac32d{\cal A}\wedge {\cal A}\wedge {\cal A}\wedge {\cal A}
  +\frac35{\cal A}\wedge
  {\cal A}\wedge {\cal A}\wedge {\cal A}\wedge {\cal A} \Big{\rangle}\;
  ,
 \eea
where $\langle\hspace{0.2em}\rangle$ denotes a cubic invariant of
the gauge algebra. Denoting the components of this invariant tensor
by $g_{\cA\cB{\cal C}}$, the field equations derived from (\ref{cs})
read
 \bea\label{cseq}
  g_{\cA\cB{\cal C}}F^{\cB}\wedge F^{{\cal C}} \ = \ 0 \;.
 \eea
Despite of its topological origin, this theory is not dynamically
trivial. For instance, in \cite{Chamseddine:1990gk} it has been
shown that specific forms of gravity in odd dimensions (with
so-called Lovelock terms) can be viewed as Chern-Simons gauge
theories based on the AdS group $SO(D-1,2)$ with cubic invariant
  \bea \la{invt}
  \langle M_{AB}M_{CD}M_{EF}\rangle\ = \ \varepsilon_{ABCDEF}\;.
 \eea
While expanding the resulting action around the maximally-symmetric
$AdS_5$ solution, i.e.~for $\langle F^{\cA}\rangle=0$, does not give
rise to a non-vanishing propagator, this is not so for generic
backgrounds with $\langle F^{\cA}\rangle\neq 0$. In fact, there is a
$AdS_4\times S^1$ solution, characterized by
 \bea\label{4Dsol}
  \bar{R}^{\alpha\beta}+\Lambda\bar{e}^{\alpha}\wedge\bar{e}^{\beta}
  \ =\ 0\;, \qquad
  \bar{R}^{\alpha 4}+\Lambda\bar{e}^{\alpha}\wedge\bar{e}^4\ \neq \
  0\;,
 \eea
(with four-dimensional indices $\alpha,\beta,\ldots=0,\ldots,3$),
which precisely propagates a \textit{four-dimensional} graviton.

Let us now turn to the construction of a Chern-Simons theory based
on a HS extension of $\mso(D-1,2)$. As these will be gauge-invariant
and extend the gravitational theory discussed above, they provide a
coupling to gravity that is consistent with the HS gauge symmetries
and thus circumvent the no-go theorems. Since the existence of
non-trivial infinite-dimensional HS algebras has already been
established, the only thing left is to define a cubic invariant of
the HS algebra, which extends (\ref{invt}). To this end we have to
introduce some mathematical machinery, notably the BCH star product.
The latter defines an associative product on the enveloping algebra
$\cU(\mso(D-1,2))$ by
 \bea \la{moyalLie}
  F(M)\star
  G(M)&=&\exp\Big(M_{AB}\L^{AB}(\del_N},{\del_{N'})\Big)F(N)G(N')\Big|_{N=M,
  N'=M }\ ,
 \eea
where $F(M)\in \cU(\mso(D-1,2))$, etc., are polynomials in the
$M_{AB}$. Here $\del_N$ is a short-hand notation for $\del/\del
N_{AB}$, and $\L^{AB}=-\L^{BA}$ is defined through the
Baker-Campell-Hausdorff relation via
 \bea
  \exp
  Q\exp{Q'}&=&\exp\big(Q+Q'+\L^{AB}(Q,Q')M_{AB}\big)\ ,
 \eea
with $Q=Q^{AB}M_{AB}$ and $Q'=Q'^{AB}M_{AB}$ for some anti-symmetric
tensors $Q^{AB}$ and $Q'^{AB}$. Next we define a sequence of traces
by insertion of a differential operator $\Delta$ that involves the
epsilon tensor,
 \begin{eqnarray} \la{grcycl}
  \Tr_k\big(F(M)\big) &=&  \tr\big(\D^k[F(M)]\big) \ \equiv \
  \D^k[F(M)]\big|_{M=0}\;, \\
  \la{Ddefn} \D &=&  \varepsilon_{A_1\cdots A_3B_1\cdots B_3}
  \frac{\del}{\del M_{A_1B_1}}\cdots \frac{\del}{\del M_{A_3
  B_3}}\;.
 \end{eqnarray}
It has been proved in \cite{Engquist:2007kz} that these traces are
\textit{cyclic}. We then define a cubic tensor as
 \bea \la{invte}
  &&\big\langle T_s,T_{s'},T_{s''}\big\rangle\ := \
  \sum_{k=1}^\infty\a_k\Tr_k\big(\{T_s,T_{s'}\}_\star \star
  T_{s''}\big)\ ,
 \eea
where $\alpha_k$ are as yet undetermined coefficients. If we view
the enveloping algebra $\cU(\mso(D-1,2))$ as a Lie algebra --- with
the star commutator as bracket ---, (\ref{invte}) is invariant under
its adjoint action. In fact, acting with an arbitrary element of
$\cU(\mso(D-1,2))$ on (\ref{invte}) yields by virtue of the
associativity of (\ref{moyalLie}) and the cyclicity of
(\ref{grcycl}) zero. Moreover, the invariant tensor (\ref{invte})
has been defined such that it reduces for the AdS subalgebra
precisely to (\ref{invt}), which fixes the first coefficient to be
$\alpha_1=\tfrac{1}{12}$. Thus the resulting HS Chern-Simons theory
provides a consistent HS extension of Lovelock gravity in $D=5$.

However, in the enveloping algebra leaving invariant (\ref{invte})
we did not yet mod out an ideal $\cI$, and so the resulting HS
theory contains various kinds of mixed Young-tableau
representations. At the same time, we have an infinite number of
coefficients $\alpha_k$ to our disposal, which might be fixed by
invariance once a more minimal choice of HS algebra has been made.
We will leave an exhaustive analysis of the possible ideals in
(\ref{hsalgebra}) and their constraints, if any, on the coefficients
$\alpha_k$ for future work.

Up to now we established the existence of HS theories that are
consistent in the sense that they exhibit by construction a HS gauge
symmetry. It remains to be shown that its dynamical content in the
free limit correctly describes massless HS modes. Put differently,
we have to show that one recovers the expected free field equations
of Fr\o nsdal \cite{Fronsdal:1978rb}. As we discussed above, for the
purely gravitational part, an expansion around $AdS_5$ does not give
rise to dynamical degrees of freedom. The same holds in presence of
HS fields. In fact, linearizing the field equations (\ref{cseq})
around a given background geometry gives
 \bea\label{CSHSeq}
  g_{\cal ABC}{\cal R}_{\rm AdS}^{\cal B}\wedge R_{\rm HS}^{\cal C}\ = \ 0\;,
 \eea
where ${\cal R}_{\rm AdS}$ denotes the AdS covariant Riemann tensor
of the background geometry and $R_{\rm HS}$ the (linearized) HS
contribution. Therefore, in case of an $AdS_5$ background with
${\cal R}_{\rm AdS} =0$ the linearized field equations are
identically satisfied and to not lead to any (perturbative)
dynamics. What we can do instead is to expand around a
non-maximally-symmetric solution as the $AdS_4$ solution in
(\ref{4Dsol}). Inserting the latter into (\ref{CSHSeq}) and assuming
vanishing HS torsion, yields precisely the HS Einstein equations for
the four-dimensional part of the HS fields. As we saw in the
previous section, even though these equations are of
higher-derivative order, they are nevertheless equivalent to the
(AdS)-Fr\o nsdal equations, or in other words, the higher
derivatives correspond to gauge degrees of freedom. Thus the
Chern-Simons action (\ref{CS}) in $D=5$ based on a HS algebra
describes in particular the dynamics of all four-dimensional
massless fields with integer spin. In this sense, it is possible to
solve the four-dimensional HS problem in a somewhat holographic way
by going to a five-dimensional topological theory!

We close with a few comments on possible impacts and generalizations
of these theories. First of all we note that the construction
directly extends to all odd dimensions \cite{Engquist:2007kz}. It
has been conjectured already some time ago that M-theory might
actually have an interpretation as a Chern-Simons gauge theory in
$D=11$ based on $OSp(1|32)$ \cite{Horava:1997dd}. Since M-theory
should contain the infinite tower of HS states described by
10-dimensional string theory, it is tempting to speculate that the
right framework might be a Chern-Simons theory based on a HS
extension of $OSp(1|32)$. The crucial test would be to see whether
such a theory permits backgrounds that are 10-dimensional and
propagate massive HS modes via some sort of spontaneous symmetry
breaking.

\section*{Acknowledgments} For useful comments and discussions we
would like to thank N.~Boulanger, D.~Francia, M.~Henneaux,
C.~Iazeolla, P.~Sundell and M.A.~Vasiliev.

This work has been supported by the European Union RTN network
MRTN-CT-2004-005104 {\it Constituents, Fundamental Forces and
Symmetries of the Universe} and the INTAS contract 03-51-6346 {\it
Strings, branes and higher-spin fields}. O.H. has been supported by
the stichting FOM.

\end{document}